\documentclass[aps,prl,twocolumn,groupedaddress, showpacs]{revtex4}
\usepackage[english]{babel}
\usepackage{amssymb,amsmath}
\usepackage{graphicx}
\usepackage{multirow}
\usepackage{bigstrut}

\begin{document}


\title{
Nanoscale superconducting gap variations, strong coupling signatures and lack of phase separation in optimally doped BaFe$_{1.86}$Co$_{0.14}$As$_{2}$
}


\author{F. Massee}\email{F.Massee@uva.nl}
\author{Y. Huang}
\author{R. Huisman}
\author{S. de Jong}
\author{J.B. Goedkoop}
\author{M.S. Golden}
\affiliation{Van der Waals-Zeeman Institute, University of Amsterdam, 1018XE Amsterdam, The Netherlands}



\date{\today}

\begin{abstract}
We present tunneling data from optimally-doped, superconducting BaFe$_{1.86}$Co$_{0.14}$As$_{2}$ and its parent compound, BaFe$_{2}$As$_{2}$. In the superconductor, clear coherence-like peaks are seen across the whole field of view, and their analysis reveals nanoscale variations in the superconducting gap value, $\Delta$.  The average magnitude of 2$\Delta$ is $\sim$7.4 k$_B$T$_c$, which exceeds the BCS weak coupling value for either s- or d-wave superconductivity. The characteristic length scales of the deviations from the average gap value, and of an anti-correlation discovered between the gap magnitude and the zero bias conductance, match well with the average separation between the Co dopant ions in the superconducting FeAs planes. The tunneling spectra themselves possess a peak-dip-hump lineshape, suggestive of a coupling of the superconducting electronic system to a well-defined bosonic mode of energy 4.7 k$_{B}T_{C}$, such as the spin resonance observed recently in inelastic neutron scattering.\end{abstract}


\maketitle


The pnictide high T$_{c}$ superconductors \cite{REF:J.Am.Chem.Soc.130(2008)} with maximal T$_{c}$'s currently exceeding 55 K \cite{REF:ChinesePhysLett25}, are the subject of global scrutiny at a level on a par with that seen for the cuprates, and more recently, graphene. One of the most debated issues is their similarities and differences with respect to the cuprates \cite{REF:Physics1(2008)}. The pnictides display many features we recognise from the cuprate repertoire, yet there are arguments that they are essentially different \cite{REF:arxiv:0808.1390v1}. In the last few years, scanning tunneling microscopy and spectroscopy (STM/STS) has played a major role in investigating the electronic structure of the cuprates on length scales down to those of the atom \cite{REF:Revmodphys79_2007, REF:Nature2003_403Davis, REF:Nature2003_422Davis, REF:Science2002_297Davis, REF:Nature2007_447Yazdani}. This effort has brought the role of intrinsic disorder introduced by dopant atoms into sharp focus. Consequently, BaFe$_{2-x}$Co$_{x}$As$_{2}$ is of great interest, not only as an electron-doped pnictide, but also because the electronically active dopants in this system are situated in the superconducting layers themselves.

\begin{figure}[h]
\includegraphics[width=8cm]{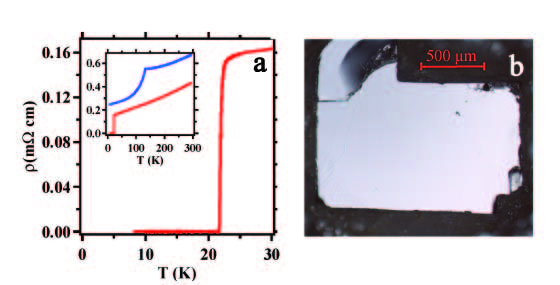}
\caption{\label{FIG:resistivity}(a) The Co-doped superconductors show a sharp transition ($\Delta$T$_{c}$ $\sim$0.5 K) into the superconducting state at 22 K (red curves).  The inset displays the full measured temperature range, together with data from the parent compound (blue trace). (b) Optical micrograph of one of the measured crystals. Very flat cleavage surfaces are typically obtained.}
\end{figure}

Single crystals of superconducting BaFe$_{1.86}$Co$_{0.14}$As$_{2}$ and the non-superconducting parent compound BaFe$_{2}$As$_{2}$ were grown using a self flux method. Typical crystal sizes are 1x1x0.1mm$^3$ (see Fig. \ref{FIG:resistivity}b). 
The high quality of our crystals is illustrated in Fig. \ref{FIG:resistivity}a, with the parent compound displaying the well-known resistivity kink at 130 K which has been associated with a SDW and accompanying orthorhombic phase transition \cite{REF:PRB78_020503(2008), REF:EPL83_27006(2008), REF:0804.3569, REF:PRL100_247002(2008)}. Co doping erases any sign of these transitions in the resistivity, but brings with it a very sharp transition into the superconducting state, which takes place at 22 K in this case.

For STM investigation, crystals were cleaved at a pressure of 5x10$^{-10}$ mbar at room temperature in a preparation chamber and immediately transferred into the STM chamber itself (base pressure $<$1.5x10$^{-11}$ mbar), where they were cooled to 4.2 K. The tunneling experiments were carried out using both electrochemically etched W and cut Pt/Ir tips, which were conditioned before each measurement on a Au(778) single crystal and yielded identical results. In all cases, the spectral shapes obtained were independent of the tip to sample distance \cite{REF:Renner95}, a sign of a good vacuum tunnel junction. Spatial drift of the STM system at 4.2 K is the equivalent of one atom diameter in 12 hours. 

Low energy electron diffraction (LEED) was performed \textit{in situ}, directly after the STM/STS measurements. For all measured samples (both pristine and Co-doped), only the tetragonal unit cell spots were seen in LEED (see Fig. 2a), with no sign of extra spots (or extinctions) as would occur as a result of a significant and structurally coherent reconstruction of the atomic positions at the surface.

We begin our discussion with the Co-doped crystals. 
A constant current image with atomic resolution is shown in Fig. 2a. 
In general, over areas  of up to 150x150 nm$^{2}$, we saw no sign of steps on the surface, with the maximal apparent corrugation being less than 1.5 \AA \ on all samples measured.
\begin{figure}[h]
\includegraphics[width=8cm]{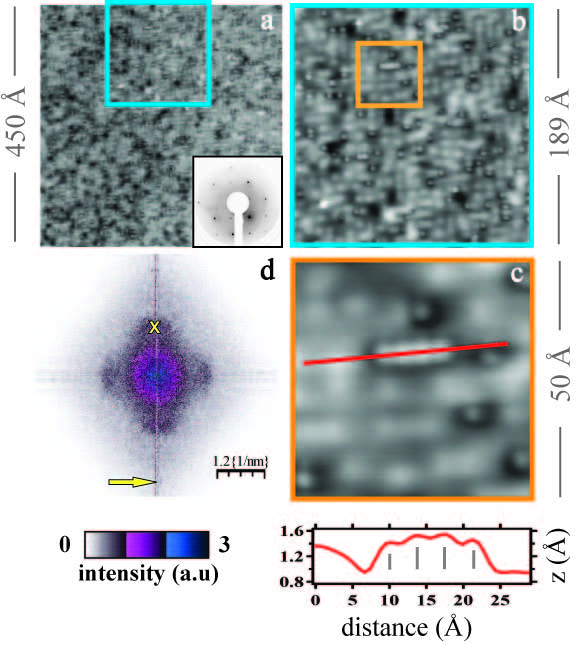}
\caption{\label{FIG:ccimages}(a) Constant current image (V$_{\text{sample}}$ = -35 mV, I$_{\text{setup}}$ = 40 pA) of Co-doped Ba122. The inset shows a LEED image from the same surface with E$_{0}$=114eV. (b) Zoom of panel (a), in which the surface atoms can be seen as bright dots.  (c) Further zoom of panel (b), showing a row of four surface atoms separated by the tetragonal cell dimension of 3.9 \AA. (d) Fourier transform of panel (a), whereby the yellow X and arrow indicate real-space distances of 8 and 3.9 \AA, respectively. All data (with exception of the LEED) were recorded at 4.2 K.}
\end{figure}
From the zooms shown in Figs. 2b and 2c, one can immediately see that the surface atoms lie arranged along the direction of the clear (1x1) tetragonal unit cell we measure using LEED, but that the inter-atomic spacing is quite irregular. The most frequent separation seen is $\sim$8 \AA: twice the tetragonal unit cell. Occasionally a row of atoms with a separation of 3.9 \AA \ occurs, as seen in panel (c), and sometimes, the (bright) atoms sit on a black background. 
The irregularity in inter-atomic distances is reflected in the Fourier transform shown in Fig. 2d, in which smeared out features predominate, corresponding to a real space separation of $\sim$8 \AA \ (marked with an 'X' on the FFT). The tetragonal lattice is barely visible in the form of weak spots, highlighted in Fig. 2d with a yellow arrow.

Inspection of the crystal structure of Ba122 leads to the supposition that cleavage occurs at the As-Ba interface. In order to avoid creation of a polar surface, the charge of the termination layer of a system like the pnictides should be -1/2 of that of the layer beneath \cite{REF:arxiv:0808.1390v1}, and this condition can be met in Ba122 if exactly half of the Ba atoms remain on each of the surfaces created by cleavage. For a room temperature cleave, the Ba ions may re-order to minimize their mutual Coulomb repulsion, resulting in inter-atomic distances larger that the in-plane tetragonal lattice constant, as seen in Fig. 2.

Although the LEED patterns from all studied surfaces show a non-reconstructed (1x1) tetragonal pattern, on an STM length scale, the details of the topography vary from cleave to cleave, and occasionally also for different locations on the same cleave. A commonly encountered variation is shown in Fig. 3a from a different Co-doped crystal of the same doping level, in which the atomic contrast is absent, and a 2D, maze-like network is seen, oriented along the tetragonal axes with typical period of $\sim$12 \AA. This image resembles previously reported one-dimensional stripe-like structures  \cite{REF:0806.4400,REF:0810.1048v1}, whereby our 'stripes' appear cut into shorter and more disordered segments, probably as a result of the higher cleavage temperature. This is in keeping with a recent report of a temperature dependence of the surface topology in a related system \cite{REF:arxiv:0812.2289v1}. 
In Fig. 3b, we show an image from pristine Ba122, which displays a very similar surface topology as in panel (a), as can also be seen from the line scans through both images shown in panel (c).

We now move on to the tunneling spectra. 
Differential conductance spectra (dI/dV) of both the Co-doped and pristine Ba122 systems were recorded (modulation amplitude of 2mV at a lock-in frequency of 427.3 Hz) on a square 64x64 pixel grid at 4.2 K.
The spectra for the Co-doped system vary significantly between different locations within a single field of view (FOV).

\begin{figure}[h]
\includegraphics[width=8cm]{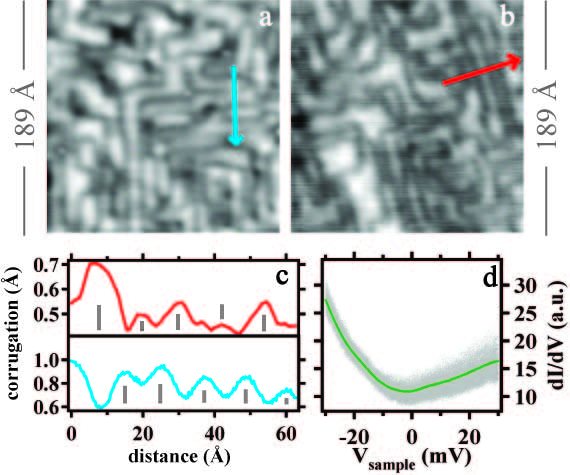}
\caption{\label{FIG:parentc}(a) Constant current image from a different sample of BaFe$_{1.86}$Co$_{0.14}$As$_{2}$ (V$_{\text{sample}}$ = -88 mV, I$_{\text{setup}}$ = 40 pA). (b) Analogous topography from BaFe$_{2}$As$_{2}$ (V$_{\text{sample}}$ = -100 mV, I$_{\text{setup}}$ = 33 pA). (c) Line scans from the  superconducting (top) and parent (bottom) compounds taken along the directions indicated by the arrows. Typical distances between the bright rows in the images is 12 \AA \, as indicated by grey ticks in the line scans. (d) Overlay of 400 differential conductance spectra from pristine Ba-122, together with their average (bold line). All data recorded at 4.2 K.}
\end{figure}

In Fig. 4a we show the result of plotting the peak-to-peak energy separation in the form of a gap map, illustrating that the major part of the FOV supports a gap, $\Delta$, of 7 meV (green areas), with smaller patches of dimension between 5 and 10 \AA \ possessing significantly smaller (red/yellow) and larger (blue) gaps. Of the 4096 spectra, there are only a handful that do not exhibit coherence-like peaks, and thus if these peaked structures represent the superconducting gap in each case, we can exclude phase separation between superconducting and non-superconducting (magnetic) regions in this optimally-doped pnictide high temperature superconductor. 
The spectra have then been binned into five groups \cite{REF:footnote_binning} and the bin averages are shown in Fig. 4c. Optimally-doped BaFe$_{2-x}$Co$_x$As$_{2}$ supports values of 2$\Delta$/k$_B$T$_c$ between 5.3 for the smaller gaps, through 7.4 for the modal gap value (7 meV) to 10.6 for the largest gaps. Our average (and most frequently occurring) gap value is close to that seen recently in ARPES measurements from optimally-doped BaFe$_{2-x}$Co$_x$As$_{2}$ for the outermost $\Gamma$centered ($\Gamma_2$ or $\beta$) Fermi surface \cite{REF:arXiv:0812.3704, REF:footnote_Gamma2}, and is practically identical to data from another STS experiment on the same material \cite{REF:0810.1048v1}. These normalised gap values are also in keeping with ARPES data \cite{REF:0808.2185v1} from the outer hole pocket Fermi surface in hole-doped (Ba,K)-122. At present it is not clear whether the tunneling matrix elements operative in STS favour the hole or electron pocket Fermi surfaces. In any case, we note that all of the normalised gap values we observe are greater than those expected for a weakly coupled BCS s- or d-wave superconductor.

\begin{figure}[h]
\includegraphics[width=8cm]{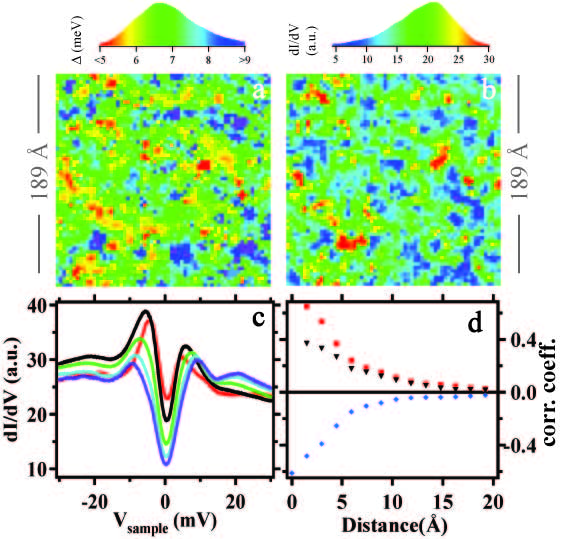}
\caption{\label{FIG:gapmap}(a) Gap map with identical FOV as Fig. 2b, taken with the same setup conditions. (b) Map of the zero bias conductance, same FOV. (c) Binned tunneling spectra for ranges of $\Delta$ (see text), plotted in colours matching those in the gap map (black has been swapped for yellow, for clarity). (d) Correlation functions. Blue diamonds show the azimuthal integrated cross-correlation between the gap and ZBC maps). The red squares and black triangles show the autocorrelation traces for the gap map shown in (a) and for an analogous data set recorded from the FOV shown in Fig. 3a respectively.}
\end{figure}

Fig. 4c also illustrates a relation between the magnitude of $\Delta$ and the zero bias conductance (ZBC), or the 'depth' of the gap. In Fig. 4b we plot a map of the ZBC, with the colour scale reversed with respect to panel (a): even a quick glance suffices to see that deeper gaps (low ZBC) mean larger gaps. This relation is formalized in panel (d) in which the azimuthal-integrated cross correlation function between the gap and ZBC maps shows an anti-correlation that dies away after some 8-10 \AA.
These spatial variations in the ZBC remind one of the impurity resonance scattering observed in the cuprates \cite{REF:Nature4_(1998)}, and may indicate the role played by the Co doping directly into the superconducting layers of this system.

Returning to the STS spectra of Fig. 4c, we point out their interesting, peak-dip-hump (PDH) line shape. For the modal gap spectra ($\Delta$=7 meV), the dip is typically located at 14 $\pm$1 meV with the hump at 20 $\pm$2 meV. If we interpret this line shape as an $\alpha^2$F fingerprint, then the peak$\leftrightarrow$hump energy separation would give a typical bosonic mode energy $\Omega_{\text{mode}}$ of 6-8 meV, corresponding to $\sim$4.7 k$_{B}$T$_{c}$.  We point out that a recently observed magnetic resonance mode seen below T$_c$ in Co-doped Ba-122 at 9.6 meV (5 k$_B$T$_c$) \cite{REF:arxiv0811.4755v1} and in Ba$_{1-x}$K$_{x}$Fe$_{2}$As$_{2}$ at ca. 14 meV (4.3 k$_B$T$_c$) \cite{REF:arxiv0807.3932} would match the peak-hump energy separation in our STS data very nicely. As an alternative, the lowest energy phonon mode observed clearly in neutron and optical measurements is at 12 meV \cite{REF:PRL101_157004(2008),REF:arxiv0806.2303v1}. Recent ARPES measurements of (Ba/Sr,K)-122 report dispersion kinks at ca. 40 meV for the $\Gamma_1$ (or $\alpha$) and 18 $\pm$5 meV for the $\Gamma_2$ (or $\beta$) Fermi surface, the latter kink being not far from the dip energy seen in the STS spectra presented here \cite{REF:footnote_Gamma2}.

The natural question upon consideration of Fig. 4 regards the origin of the observed spatial disorder in gap values.  Firstly, how about the non-trivial surface topography illustrated in Figs. 2 and 3 ? 
Pristine Ba122 offers the best test of this point, as it is without substitutional disorder within the FeAs structural unit. Fig. 3d shows a compilation of one tenth of a complete STS data set (400 of 4096 STS spectra) - and their average - taken across the same FOV as Fig. 3b. It is immediately clear that the near-$E_{F}$ electronic states in pristine Ba-122 are highly spatially homogeneous, both in terms of the shape, as well as in terms of the absolute dI/dV value. Thus, the gap disorder seen in Fig. 4 is a property of the superconducting system, and thus the finger is quickly pointed in the direction of the Co dopants as the culprit.
For our doping level, 1 in 14 Fe atoms is replaced by a Co dopant, and in the simplest of pictures this gives a Co - Co separation of a little over 10 \AA. If the cobalt ions were to be disordered - i.e. also appearing, locally, at higher and lower concentrations than 14\% - the length scale of these departures would be this same 10 \AA.
In Fig. 4d we show autocorrelation traces for the gap map [panel (a)] and the same quantity for a gap map recorded from an identical FOV as Fig. 3a. Both these (positive) correlation traces show that 8 \AA \ is the characteristic length scale of the significant gap variations away for the 'background' value of 7 meV, which is very close to the Co - Co length scale. The fact that the atomically-resolved cleaves (Fig. 2a) and those with the 2D maze-like structure (Fig. 3a) both give the same characteristic length scales for the deviations from the average gap (see Fig. 4d), is a further indication that the topographic details - which most likely track the particulars of the surface Ba (dis)order - do not have much direct effect on the superconducting system.

How do the spatial gap variations found here in an optimally (electron) doped pnictide compare with those we know from STS studies of the cuprate high T$_c$'s at analogous doping levels ? Optimally doped Bi2212 yields a similar total spread of a factor two in normalised gap values, but upshifted with respect to those here to lie between 6 and 13 k$_B$T$_c$ \cite{REF:Nature2003_422Davis}. Recently, the emphasis has come to lie on the role played by the pseudogap in the observed large apparent superconducting gap disorder seen in the cuprates \cite{REF:Boyer_HudsonNatPhys3_802_2007}. In the pnictide STS data presented here, it would be natural to take the modal gap value as that representing areas with Co doping occurring in the FeAs plane at the nominal level. If we reason that Co clustering leading to a local overdoping would lead to a local reduction in $\Delta$, then a process of elimination would link the large gaps observed to Co-deficient - underdoped - regions of the sample. There have been numerous indications of pseudogap behavior in the pnictides, and we cannot conclude at present whether the significant nanoscale inhomogeneity in the superconducting gap value is wholly unrelated to pseudogap effects. Detailed temperature and doping dependent measurements are required for a resolution of this issue.

In summary, we present detailed STM and STS investigations of pristine Ba122 and samples of the electron doped pnictide superconductor BaFe$_{1.86}$Co$_{0.14}$As$_{2}$, which is close to optimal doping. In the first part of the paper we describe the complex topography of the surfaces of these single crystals, which is probably a result of partial lift-off of the Ba ions upon cleavage. We go on to demonstrate that this termination-plane topographic disorder has little effect on the low lying electronic states of these systems. 

The STS data from the superconducting samples display clear coherence-peak-like features, demarcating an energy gap which, on average, is of magnitude 7.4 k$_B$T$_c$. There exist, however, significant spatial deviations from this modal gap value, with the gap distribution spanning a range from 5 to 10 k$_B$T$_c$. If these gaps are indeed superconducting gaps, we can clearly rule out nanoscopic phase separation in these samples.
There is a robust anti-correlation between the peak-to-peak separation and the zero bias conductance, which operates over length scales of order 8 \AA. The same length scale reappears upon an analysis of the spatial correlation of the low and high gap deviations from the background gap value. All these length scales are very close to the average separation of the Co atoms in the FeAs superconducting blocks, thus highlighting their importance as local dopants, and effective scatterers/pair-breakers for the superconducting condensate in this material.  

Finally, the STS spectra themselves suggest significant structure in the imaginary part of the electronic self energy of this system at energies less than 10 meV away from the gap edges, which makes itself felt as a peak-dip-hump line shape. This energy scale (normalised to $T_{C}$) matches that of a magnetic resonance mode seen in inelastic neutron scattering of the both the hole and electron doped pnictide high T$_c$ superconductors.

\begin{acknowledgments}
We thank H. Luigjes, H. Schlatter and J. S. Agema for valuable technical support, and J. van den Brink and A. de Visser for useful discussions. This work is part of the research programme of the FOM, which is financially supported by the NWO, and was also supported by the University of Amsterdam.
\end{acknowledgments}


\begin{thebibliography}{99}
\bibitem{REF:footnote_binning}
		The STS data are binned into five sets with the following $\Delta$ ranges: 0 to 5.5 meV, 5.5 to 6.5 meV, 6.5 to 7.5 meV, 7.5 to 8.5 meV, 8.5 to $\Delta_{max}$ meV.
\end{thebibliography}

\end{document}